\documentclass[10pt,conference]{IEEEtran}
\IEEEoverridecommandlockouts
\usepackage{amsmath,amssymb,amsfonts}
\usepackage{import}
\usepackage{graphicx}
\usepackage{subcaption}
\usepackage{textcomp}
\usepackage{xcolor}
\usepackage{url}
\usepackage{listings}
\usepackage{xspace}
\usepackage{multirow}
\usepackage{colortbl}
\usepackage{blindtext}
\usepackage{float}
\usepackage{ifthen}
\usepackage{rotating}
\newboolean{showcomments}
\setboolean{showcomments}{true}
\ifthenelse{\boolean{showcomments}}
  {\newcommand{\bnote}[2]{
	\fbox{\bfseries\sffamily\scriptsize#1}
    {\sf\small$\blacktriangleright$\textit{#2}$\blacktriangleleft$}
   }
   \newcommand{\paragraphDesc}[2]{
    \textcolor{#1}{#2}
   }
   
  }
  {\newcommand{\bnote}[2]{}
   
   \newcommand{\paragraphDesc}[2]{}
  }

\graphicspath{{figures/}}

\newcommand{\figref}[1]{Figure~\ref{fig:#1}}

\newcommand{\tabref}[1]{Table~\ref{tab:#1}}

\newcommand{\commented}[1]{}

\newcommand{\ie}{\emph{i.e.,}\xspace}

\newcommand{\etc}[1]{\textit{etc.}}

\usepackage{url}            
\makeatletter
\def\url@leostyle{%
  \@ifundefined{selectfont}{\def\UrlFont{\sf}}{\def\UrlFont{\small\sffamily}}}
\makeatother
\usepackage{pifont}
\newcommand{\cmark}{\ding{51}}%

\newcommand{\productOwner}{Product Owner\xspace}

\usepackage[scaled=0.85]{helvet}

\begin{document}

\title{Report From The Trenches \\
\huge{A Case Study In Modernizing Software Development Practices}}

\author{\IEEEauthorblockN{Houékpétodji Mahugnon Honoré\IEEEauthorrefmark{1}\IEEEauthorrefmark{3},
Nicolas Anquetil\IEEEauthorrefmark{2}\IEEEauthorrefmark{3}, Stéphane Ducasse\IEEEauthorrefmark{3},\\
Fatiha Djareddir\IEEEauthorrefmark{1}, Jérôme Sudich\IEEEauthorrefmark{1}}
\IEEEauthorblockA{\IEEEauthorrefmark{1}SA-CIM, Univ. Lille, CNRS, Inria, Centrale Lille, UMR 9189 CRIStAL, 59000 Lille, France}
\IEEEauthorblockA{\IEEEauthorrefmark{2}Université de Lille, CNRS, Inria, Centrale Lille,UMR 9189  CRIStAL France,ORCID: 0000-0003-1486-8399}
\IEEEauthorblockA{\IEEEauthorrefmark{3}Université de Lille, CNRS, Inria, Centrale Lille, UMR 9189  CRIStAL France,ORCID: 0000-0001-6070-6599}

homahugnon@gmail.com, nicolas.anquetil@univ-lille.fr, stephane.ducasse@inria.fr}

\maketitle

\begin{abstract}
One factor of success in software development companies is their ability to deliver good quality products, fast.
For this, they need to improve their software development practices.
We work with a medium-sized company modernizing its development practices.
The company introduced several practices recommended in agile development.
If the benefits of these practices are well documented, the impact of such changes on the developers is less well known.
We follow this modernization before and during the COVID-19 outbreak.
This paper presents an empirical study of the perceived benefit and drawback of these practices as well as the impact of COVID-19 on the company's employees.
One of the conclusions, is the additional difficulties created by obsolete technologies to adapt the technology itself and the development practices it encourages to modern standards.
\end{abstract}


\section{Introduction}
\label{sec:introduction}

Nowadays, software is developed in a demanding environment.
The requirements change as business needs change; time to market, budget, quality, or maintainability concerns increase the level of challenge faced by software companies. 
To cope with these challenges, companies turn to modern software development processes such as agile development \cite{Bec04,Schw04}.
These processes are known to focus on managing time to market constraints and  the ability to accommodate changes during the software development life cycle \cite{Cao09}. 
They also promote better communication between the various stakeholders of a project.

If the benefits of such development processes are well documented, the transition from old practices also presents challenges \cite{Chen15}.
We work with a medium-sized company trying to modernize its development practices.
The company introduced several practices recommended in agile development, hoping to improve its product quality and time to market.
At the same time, it faced the exceptional conditions linked to the COVID-19 crisis with the generalization of remote working. 
Many agilists advocate the importance of collocation, face-to-face interaction, and physical artefacts incorporated in the shared workspace, which the COVID-19 pandemic made unavailable \cite{Smit21}.

The aim of this work is to evaluate if all the practices introduced bring the benefits expected, and how they are perceived by the various stakeholders in the company.
We aim to identify the perceived benefits of these changes, and the challenges professionals face with the practices' modernization.
We also evaluate if the context of remote working due to COVID-19 had an influence on these practices.

We conducted an exploratory study by leveraging in depth, semi-structured interviews with 17  participants in the development process to understand how these changes impacted their work.
We adopted the grounded theory approach \cite{Khad14}, a qualitative technique, to analyse interviews.
When possible, the findings of the interviews were validated with empirical data (company internal issue database, source code management commit history, emails).

Our study makes the following contributions:
\begin{enumerate}
    \item We document the industrial perception of some agile development practices;
    \item We document perceived benefits and drawbacks of these practices;
    \item We identify the impact of remote working on some agile practices.
\end{enumerate}

In Section \ref{sec:context}, we describe
the context of this study.
In Section \ref{sec:bckground}, we present the company and software development practices before modernization.
In Section \ref{sec:changed-practice} we detail changes  in practices introduced in the company.
In Section \ref{sec:study-design}  we describe our study methodology.
 Section \ref{sec:result} presents the results and Section \ref{sec:threats} discuss the threats to validity.
Finally,  in Section \ref{sec:conclusion} we conclude our research.

\section{Agile Methodologies}
\label{sec:context}

According to \cite{Vohr13}, an iterative software development process is a process in which the software is built in several successive sprints \cite{Larm04}.
Each sprint is viewed as a mini project including requirement analysis, programming and testing.
At the end of the sprint, a stable version of the project is released. 

SCRUM appears to be the most popular agile methodology  in the industry \cite{Hans18}.
It recommends sprints of 2 to 4 weeks \cite{Schw04}.
It has 3 primary roles:  Product Owner, SCRUM Master, and team member.  
The lifecycle is composed of 4 meetings:  Sprint planning, Daily SCRUM, Sprint Review, and retrospective meeting \cite{Maha13}. 

SCRUM implies tight communication (both formal and informal) within a fully dedicated project team \cite{Sadu10}.
Tools play an important role in  tracking project progress, facilitating communication and making decisions \cite{Neru05}.
These tools are also essential for source code management or code review.

\section{Practices in Industrial Case Study}
\label{sec:bckground}

This research, is a case study  on software practices modernization in a medium-sized company.
It is important to characterize precisely the context of a case study so that people can understand whether the finding may apply elsewhere.
Here we present the industrial background of this study. 

The leading software system of the company 
 is over 20 years old.
It is written in PowerBuilder\footnote{\url{https://www.appeon.com/products/powerbuilder}} programming language. A so-called ``Fourth Generation programming Language''\footnote{\url{https://en.wikipedia.org/wiki/Fourth-generation_programming_language}}, including a programming language, an Integrated Development Environment, database management, report generation, and Graphical User Interface development.
The system has 3 MLOC, in 117 PowerBuilder libraries (packages).
The largest library is over 300 KLOC.

Some difficulties in the evolution of the system started to appear, with increasing time to close issues and difficulty to deliver new features to the clients in time.
It was decided that a change in development practices needed to be introduced.

In this section, we describe the state of practice in the company before the introduction of changes.
We divided the practices into six categories:  stand-up meeting, team organization, development life cycle,  code review, code quality, issue  management.
We will discuss each of these in the following.

\subsection{Stand-up meeting}
\label{sec:bck-standup}

The company used some kind of stand-up meetings for the programming team.
These meetings occurred twice a week (Tuesdays and Thursdays).
They took place in the morning.
Each programmer in turn would explain what he had done the previous day, the problems encountered, and what he was planning to do the following day.

These meetings allowed programmers to seek help from their colleagues.
But on the other hand, such help could turn the stand-up meetings very lengthy as a particular issue was discussed by two colleagues, leaving the others to wait.

\subsection{Team organization}
\label{sec:bck-team}

This study focuses on the people involved in the development of the main product of the company, this includes a customer management team, an analysis team, a programming team, and a testing team.

The \textit{customer management team} was in direct contact with the clients, it received bug reports and enhancement request from them.
It formalized these demands in issues that are called ``tickets'' in the company.
It prioritized the tickets and followed their development.
Once a ticket was closed, the team ensured the delivery of the product.

\textit{The analysis team} totaled 4 analysts and was responsible for analyzing tickets and writing specifications (functional or non-functional) for the most complex ones (bug reports were rarely analyzed by this team).
The resulting specification was formalized in a document attached to the ticket.


\textit{The programming team} was product-oriented and totaled 15 programmers.
They were responsible for implementing tickets on a given product.

Finally, \textit{the testing team} performed functional and non-functional tests on the modified products. 
It included 8 testers.
They performed functional tests on a product and also specifically checked the tickets implemented against the specifications from the analysis team.

This organization was the source of several difficulties:
\begin{itemize}
\item Programmers were reluctant to schedule meetings with the analysis team to ask for more information (lacking in the specification document) because of delays it introduced.
As a result, the solution proposed was not always conform with the client expectations.

\item The same issue occurred with the testing team that was working on the ticket specification, thus sometimes testers were unable to check what the client required.
This would result in corrections to the ticket requested from the client and a lot of rework.

\item Such corrections of a ticket by a client after delivery meant coming back to a work done several weeks or months before.

\item Lack of communication between the teams caused tensions and instances of the ``blame game''.

\item Tickets were estimated and assigned to specific programmers by the development manager.
It came up in our interviews that, as a result, programmers felt they were dealing with one ticket after the other without a sense of an overall direction or why the tickets were important.

\item As new programmers were hired, it was more and more difficult for the team manager to oversee the work of all the members.

\item Tickets had to be studied and re-scheduled at various stages, first the analysis team, then the development team, and the testing team. 
This resulted in delays (because each team had its priorities) and work duplication.
\end{itemize}

\subsection{Software development cycle}
\label{sec:bck-sprint}

Development in the company was organized in cycles of approximately three months.
At the beginning of the cycle, a set of tickets was selected to be resolved according to their priorities and age.
The tickets were assigned to available programmers.
At the end of each cycle, one major version would be issued.

During the cycle, some urgent tickets (typically bugs, but also small evolutions less than three man-days) would appear, requiring a programmer to set aside his work to solve it.
As a consequence the planning was not respected, cycles did not deliver all that was expected, causing frustration to the clients and developers (who had to shift their focus from one ticket to another).
This would also introduce delays between a ticket specification (analysis team), its implementation (development team), and its testing (testing team).

\subsection{Source code management}
\label{sec:bck-svn}

PowerBuilder imposes to store source code in a proprietary format not easily handled by external tools.
It supports version control systems such as SVN or GIT only since 2017.
For these reasons, version control tools were never used on the system studied.
The source code was managed manually.
The product was versioned in patch-versions, major-versions, and some client-specific versions.
All versions were archived in directories on the company intranet.
There were a bit more than 1000 versions when the study started.
Versions before 2012 are completely lost.

Such practice resulted in several problems:
\begin{itemize}
\item Programmers relied on informal communication to avoid working at the same time on the same parts of the code.

\item Merging modifications was completely manual, often using a generic file comparison tool like WinMerge.

\item Changes were integrated into the ``repository'' by manually copying code from the programmer version to the ``repository'' version.

\item When doing a change on several parts of the product, programmers needed to keep a list of all impacted parts to port them in the ``repository'' one by one later.
Forgetting one part would mean, the change was not completely replicated in the ``repository'' and therefore the final version was not correct.

\item After a source code modification, the programmer had to tag the modification with a special ``property'' in PowerBuilder.
Failing to do so could mean that somebody else, unaware of the modification, would copy old source code over the (untagged) new code in the ``repository''.

\item  It was common practice to put a comment in the code to describe the modification and the date and programmer who did it.
This could help debugging the code later.
\end{itemize}

\subsection{Code quality}
\label{sec:bck-quality}

Periodic \textit{team code review} meetings (in the development teams) were organized to try to promote better code quality.
They occurred every two weeks and lasted about one hour.
Programmers were asked to prepare code snippets demonstrating bad or good coding practices.
They would then be discussed to ensure homogeneity of coding.

The subjects discussed depended very much on what the programmers had found and there would sometimes be very little to discuss in a meeting, either due to lack of time, or because programmers were not sure what would constitute a ``new and interesting'' practice for their colleagues.

These meetings were often an occasion to remind all programmers of some basic programming rules in the company: naming conventions for variables, use of a special ``infinity'' constant, {\ldots}
However, many violations of these rules could be found in the code.
Programmers were supposed to correct them when they found such violations, but it was rarely done for lack of time.

Another action promoting code quality was set up for young programmers:\textit{ individual code review}.
They were systematic when a new programmer joined the team, or when a seasoned programmer worked on a new domain for the first time.

It must be noted that no unit testing tool or practice was in place.
Testing was performed manually by each programmer on the code  just developed.

\subsection{Issue management}
\label{sec:bck-ticket}

To perform any change on the system, a \textit{ticket} is created.
It represents a unit of work.
Tickets are stored in the tickets' database since 2000.
The ticket database drives the entire software process: assignment of work to programmers, management of the workflow to answer a client request, billing information about each task.
There are tickets for fixing defects, writing documentation, adding new features, etc. 

Some data stored on tickets are:
creation date; closing date; time estimation to work on the ticket; time actually spent on the ticket (to analyse it, to implement a solution, to test the solution); programmer in charge, whether there was a follow-up ticket, etc.

\section{Practices Modernization Actions}
\label{sec:changed-practice}

The management in the company understood that the situation was sub-optimal.
Mainly, there were concerns about the rhythm of deliveries (too slow), the predictability of new features delivered, the adequacy of the delivery with the clients' needs, and the general quality of the code.

To improve the situation and resolve some of these issues, in 2019 until now, the company started to introduce changes in its practices.
These actions tackle different aspects from human resource to source code management: (1) daily stand-up meetings, (2) development team re-organization, (3) better-organized software development cycles, (4) source code management, (5) more systematic checking of code quality, and (6) dashboard of tickets.




\subsection{Stand-up meetings}
\label{sec:chg-standup}

At the beginning of 2020, the pre-existing stand-up meeting practice was made more in accordance to recommendations in agile development with daily meetings instead of twice a week.
It happens at the beginning of the day.
Participants are not only programmers (as before), but also some testers and requirement analysts (see Section \ref{sec:chg-team}).
The meeting is re-focused to treat only the three questions (for everyone): what was done the previous day; what problems were encountered; and what were the plans for the present day.
To keep the meeting short (15 to 20 minutes), if a participant has an issue that someone could help him solve, both are encouraged to discuss it after the meeting.

The CTO would like to go further by relating the stand-up meetings (micro perspective) to the objectives of the development cycle (broader perspective, Section \ref{sec:chg-sprint}) thus being able to check daily whether the team is still on track to realize this objective. This was not implemented yet.


\subsection{Team organizational change} 
\label{sec:chg-team}

A fundamental change was in the team structuring.
This is a normal consequence of modernizing software development practices as stated by \cite{Bass01a}.
In the end of 2020, the company reorganized the developers according to business domains, and an additional team was created (the ``run team'') to handle small urgent issues.

As proposed in \cite{Dorn18}, teams are reorganized so that all stakeholders in the company for a given business line are grouped in one team including programmers, a \productOwner, and a product QA.
Two such teams were created.

\paragraph*{\productOwner} the \productOwner  responsibility is to analyse and dispatch tickets to programmers.
This includes making sure that developments meet the clients' requirements.

\paragraph*{Product QA}The product QA main task is to perform functional testing on new developments.
He validates the developments or reports regressions.
This approach is qualified as  "you build it, you run it" \cite{Dorn18}.

This reorganization allows programmers to be better integrated to the source of evolutions (the \productOwner) and the tester.
It avoids misunderstandings and blame gaming.

\paragraph*{``\textit{Run team}''} This third team was created to handle urgent demands.
It handles only demands up to two man-days work (bug fixing or evolution).
This ensures that such demands do not break the workflow of the other teams.
To speed up things, demands are not specified formally, but interactively by a programmer and a customer manager.

The ``run team'' is composed of programmers, customer managers, and a technical expert.
Programmers rotate between the ``run team'' and the two other teams.
At first, they rotated at each development cycle (see Section \ref{sec:chg-sprint}), but this made it difficult to plan developments.
Now, programmers rotate every three development cycles.

\subsection{Software development cycle}
\label{sec:chg-sprint}

As explained in Section \ref{sec:bck-sprint}, the company used development cycles, but they were not formalized as recommended by agile approaches.
The main change introduced, at the begining of 2020, was to have shorter cycles, of two weeks (10 working days) instead of three months.
This allowed the company to better plan the cycles: 8 days are devoted to the development and 2 days are kept to deal with the unexpected.


On the eve of each cycle, a \textit{planning meeting}, lasting about 1.5 hour, decides what tickets should be considered for this cycle.
Participants in the decision are the \productOwner (for a given team), the customer manager, CTO  and the development manager.
The decision is based on tickets priority, and estimated working time.

Starting the cycle, there is a launch meeting, where the tickets are discussed in the development team and assigned to programmers.
At the end of this launch meeting, each programmer has a schedule of tasks for the cycle.


Closing a cycle, there is a retrospective meeting with the development team, the CTO and the development manager.
The meeting lasts one hour and analyses the problems and delays of the cycle and proposes solutions to them.
The team reorganization described in Section \ref{sec:chg-team} actually came as a result of these retrospective meetings.

Differently from recommended agile practices, releases do not occur at the end of each development cycle but keep the old schedule: patch-release every week and ``major'' release every three months.

\subsection{Source code management}
\label{sec:chg-svn}

Another step in modernizing practices in the company was the introduction of a version control system at the end of 2019.
This introduction faced two major challenges.
The first one is due to PowerBuilder, the programming language environment.
As explained in Section \ref{sec:bck-svn}, this technology did not have software version management until relatively recently.
And this management still has some quirks.
The second challenge was that, as a natural consequence, programmers were not familiar with the practice itself.

With these constraints, the company elected to introduce the Subversion (SVN) version control system in 2019.
For programmers who had no previous knowledge of version control management, SVN, with its central repository and local working copies, was deemed easier to understand than more recent (decentralised) tools like Git.
Otte \cite{Otte09} does conclude that SVN has a better user interface.

The company organised 8 hours of training for the development team.
Introducing SVN was not without difficulties:
\begin{itemize}
\item As must be expected, programmers were wary and reluctant of changing their working habits.
The Programmer 5 that had previous knowledge of SVN, acted as a champion of the technology by helping his colleagues;

\item Powerbuilder stores the source code in a binary format (PBL files) whereas version control tools need text.
This forces programmers to commit both the binary (PBL) and textual forms of the project in SVN.
The binary form is needed to be able to re-import a version in the IDE. The textual form is needed for merging capabilities.
\end{itemize}

After 18 months of use there are 4639 commits in the SVN repository and programmers are now convinced of its utility.

\subsection{Code quality}
\label{sec:chg-quality}

Two actions were set up to promote code quality: manual code reviews (see Section \ref{sec:bck-quality}), and automated code reviews (static code analysis tool or ``linter'').

Automated unit testing is seen as an other important practice to introduce, but again the PowerBuilder technology creates some roadblocks for this. 
To prepare programmers for this next disrupting change, a linter (static code analysis tool) was designed, in the begining of 2021, to check some simple code quality rules.
The idea was to accustom programmers to have an automated tool checking their commits and warning them of potential problems in their code.
The hope is that they will be more willing to adopt a testing tool and the changes it will impose in their programming habit, once they understand better the benefit they could get from it.
At the same time, a linter will also allow improving the source code progressively, something that the existing manual code reviews have difficulties to ensure.

At first, a simple rule checker was implemented with nine in-house rules: five rules checking variable naming conventions, rules on the maximum number of parameters of functions, mandatory \texttt{ElseCase} in all \texttt{ChooseCase} (i.e. \texttt{switch}) etc.
This simple tool checks each commit and sends an email to the committer for every new violation introduced or every old violation removed.


\subsection{Dashboard of tickets}
\label{sec:chg-ticket}

To try to better monitor the development activity and the possible gains resulting from the new practices, an analysis of the tickets' database is computed monthly and displayed in a dashboard since the outset of the changes (2019).

The analysis collects information on:
\begin{itemize}
\item average  time between opening and closing dates;
\item average working hours spent on a ticket;
\item average testing hours for programmers spent on a ticket; 
\item average testing hours for testers spent on a ticket;
\item average difference between time estimated and spent on a ticket. 
\end{itemize} 

\begin{figure}
    \centering
     \begin{subfigure}[b]{0.45\textwidth}
         \centering
    \includegraphics[width=\textwidth]{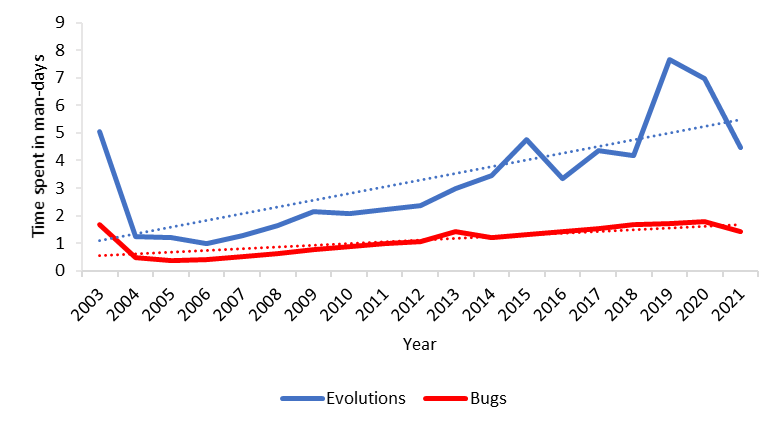}
   \caption{programmer work time on closed tickets}
         \label{fig:time_spent}
    \end{subfigure}

    \begin{subfigure}[b]{0.45\textwidth}
         \centering 
         \includegraphics[width=\textwidth]{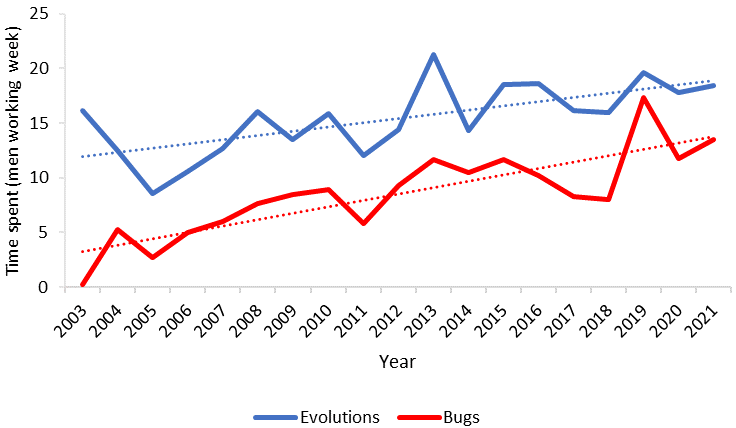}
          \caption{Lifetime of closed tickets (from opening date to closing date)}
    \label{fig:time_close}
  \end{subfigure}
   \caption{Tickets Dashboard}
    \label{fig:dashboard}
\end{figure}

The results of the analysis are presented in several graphs on a web page on the intranet.
\figref{dashboard} presents some graphs of the dashboard. 
Plots in red represent bug tickets (fix, regression, etc.), plots in blue represent evolution tickets (new features).
\figref{time_spent} shows average development time spent on closed tickets, 
\figref{time_close} shows average time for closing tickets.
These figures will be commented in Section \ref{sec:result-team}, they are presented here as examples.
This dashboard is updated monthly, allowing to track the system changes over time.

\subsection{COVID}
\label{sec:chg-covid}

As everywhere else, the COVID-19 pandemic imposed drastic working changes on the company.
Developers are encouraged to work remotely, and most communication occurs through video conferencing.
This had consequences on the practices described above.
An important one being that the stand-up meetings that were held in-person have switched to remote meetings (see Section \ref{sec:result-team}).

\section{Study Design}
\label{sec:study-design}

This paper presents an empirical study of modernizing software development practices in a medium-sized company.
The study is conducted using two different research techniques: interviews of stakeholders and data collection from the ticket database or the commit history.

\begin{table}[htbp]
    \centering 
\caption{ List of people affected by practices modernization. Total work experience an seniority in the company in years. Last column indicate participants in the interviews.  }
\label{tab:participants}
\begin{tabular}{llcc}
    \hline
    Team members & Experience & Seniority & \\
    \hline
Programmer 1 & 1 y.  of Powerbuilder & 1 y. &\cmark  \\
Programmer 2 &  12 y. of PowerBuilder & 7 y. &\cmark\\
Programmer 3 & 6 y.  of Powerbuilder& 6 y. & \cmark \\
Programmer 4 & 10 y. of Powerbuilder & 10 y. &\cmark\\
Programmer 5 &  20 y. including Powerbuilder & 3 y. &\cmark\\
Programmer 6 & 22  y. of Powerbuilder & 22 y. &\cmark\\
Programmer 7 & 20 y. of Cobol \& PLSQL  & 2.5 y. &\cmark\\
Programmer 8 & 1 y. of PowerBuilder & 1 y. &\cmark\\
Programmer 9 & 2 y. of PowerBuilder & 2 y. & - \\
Programmer 10 & 1.5 y. of Powerbuilder & 1.5 y. &\cmark \\
Programmer 11 & 7 y.  of PowerBuilder & 7 y. &\cmark\\
Programmer 12 & 11 y. & 11 y. & - \\

Dev. manager & 7 y. as development manager & 18 y.  & \cmark \\
CTO & 19 y. of project management & 2 y. & \cmark \\
                     
Tester 1  &  1 y.   & 1 y. &\cmark\\
Tester 2  &  9 y.  & 9 y. & - \\
Tester 3  &  15 y. & 7 y. & - \\

\multirow{2}{*}{Testing manager} & 13 y. analysis \& developments & \multirow{2}{*}{3 y.}  & \multirow{2}{*}{\cmark}  \\
                                 & 7 y. testing manager &  &\\

Analyst 1 & 17 y. analysis \& developments &  9 y. &\cmark \\
Analyst 2 & 23 y. analysis \& developments & 3 y.  &\cmark\\
Analyst 3 & 7 y.   & 6 y.  & - \\
Analyst 4 & 11 y. & 11 y.  & - \\

Customer mangr 1 & 17 y. & 17 y. &\cmark\\
Customer mangr 2 & 9 y.  & 2.5 y. & - \\
Customer mangr 3 & 2 y.  & 2 y. & - \\
\hline
\end{tabular}
\end{table}

\begin{table*}[htbp]
    \centering

    \newlength{\firstCol}
    \settowidth{\firstCol}{Technical}
    \newlength{\secondCol}
    \setlength{\secondCol}{3cm}

    \caption{ List of interview questions}
    \label{tab:questions}
    \begin{tabular}{lp{\secondCol}p{3.2cm}l}
    \hline
         Categories                     &Members                                  & Practices                                                        & Questions \\
    \hline     
        \multirow{5}{*}{\parbox{\firstCol}{Non-Technical}}
        & A customer manager
        & Software development cycle
        & Are you aware of the introduction of this practice? \\
        
        & A Tester
        & Team organization
        & What does this practice consist of? \\
        
        & The testing manager
        & Dashboard of tickets
        & What was the situation before the introduction of this practice? \\                                              
        & Analysts
        &
        & What is the situation now with the practice in place? \\
        
        &
        &
        & What improvement can you propose to this practice? \\
    \hline
        \multirow{7}{*}{Technical}
        & Developers
        & Stand-up meeting
        & Are you aware of the introduction of this practice? \\
        
        & The development  manager
        & Team organization
        & What does this practice consist of? \\
        
        &
        & Software development cycle 
        & What was the situation before the introduction of this practice? \\
        
        &
        &  Source code management
        & What is the situation now with the practice in place? \\
        
        &
        & Dashboard of tickets
        & What improvement can you propose to this practice? \\
        
        &
        & Code review
        & \\
        
        &
        & Linter
        & \\
                                      
    \hline
        \multirow{9}{*}{CTO}
        &
        & Stand-up meeting
        & What does this practice consist of?\\
        
        &
        & Team organization
        &  What was the situation before the introduction of this practice?\\
        
        &
        & Software development cycle 
        & Why was it a problem? \\
        
        &
        & Source code management 
        & How was the practice chosen? \\
        
        &
        & Dashboard of tickets
        &  What were the implementation difficulties?\\
        
        &
        &  Code review
        &  What is the situation now with the practice in place?\\
        
        &
        & Linter
        & What changes did it bring  better/less good? \\
        
        &
        & 
        &   What improvement can you propose to this practice?\\

    \hline
    \end{tabular}
    \end{table*}

We conducted semi-structured in-depth interviews to identify how software development practices and their modernizing are perceived by the people in the company.
We conducted 17 interviews with people of varied responsibilities:
\begin{itemize}
    \item Ten software programmers;
    \item A tester;
    \item The testing manager; 
    \item A customer manager;
    \item Two analysts who are also Product Owners;
    \item The development manager;
    \item The CTO.
\end{itemize}

\tabref{participants} gives some data on all people impacted by the practices modernization:
their overall experience (in software development related activities), their seniority in the company, and whether they participated in the interviews.

Due to the COVID-19 pandemic and remote working, some interviews were performed remotely.
All interviews were recorded to ease analysis and we asked consent of the participants for that.
We also reassured participants on the anonymity of their answers.
Note that to promote this anonymity, all interviewees are refereed to as ``he'' in the paper, even women.
Depending on participants, interviews lasted from 30 minutes to 2 hours. 
The interviews followed a questionnaire were five questions were asked for each of the practice considered.
\tabref{questions} shows the questions and the practice considered for each 
category of participants: CTO, technical personnel (programmers and the development  manager), and non-technical personnel (all other personnel).
The CTO questionnaire was different because, first, he is the initiator of many of the practice changes (he knows the why) and, second, he is less directly impacted by the changes since he does not take part in the concrete development activities (analysis, coding, testing)
Interviewees were also asked how COVID-19 pandemic and remote working impacted their work.

For analyzing the 17 semi-structured interviews, we followed the Grounded Theory approach, a qualitative technique, inspired by \cite{Khad14}.
Grounded Theory is an exploratory research method that aims at discovering new perspectives and insights, rather than confirming existing ones \cite{Char14a}.

After recording, interviews were transcribed and anonymized.
The transcripts were analyzed, with MAXQDA \cite{Kuck19}, through \textit{coding} (or labeling).
Each transcript was broken into a code database by assigning a label to the main idea of sentences until saturation.
Next we iteratively grouped \textit{codes} into \textit{concepts} which are then organized by interview questions.

Apart from the interviews, we also used factual data from the tickets' database, and commit history, to corroborate the perception of the participants.

In the next section, we present the results of our study.

\section{Results}
\label{sec:result}

In this section, we present the interviewees' point of view on the modernization actions initiated by the company.

\subsection{Stand-up meeting}
\label{sec:result-standup}

With meetings every other day (as it was before the changes), there was less pressure to analyze the root cause for delays.
People used to answer the question ``What did I do?'', but did not always reflect on ``What held me back?''.
Having the stand-up meetings every day forces to self-reflect on the issues and their solutions.
The Programmer 3 reports that ``\emph{Before, there was less pressure. We were supposed to say what we were working on, and that was it.
But now everyone is forced to think about what him back. Because every day he has to analyze his issues}''

Some view it as a good way to organize their work (or their day).
It also creates an extra motivation to finish during the day what was planned in the morning.
This is a known advantage of the daily meetings that meetings every two days didn't seem to bring.

The meetings are now more focused, the Programmer 5 states that they are no longer trying to solve technical problems during the meeting, but focus on the three questions ``What did I do?'', ``What held me back?'', ``What will I do today?''.
Yet, several note that the meetings don't always end after the planned 15 to 20 minutes and are still not focused enough.
Another regrets that remarks are not always constructive.

In the old version, people tended to help each other during the stand-up meeting, thus the time of the entire team was ``wasted'' while two people might discuss a particular issue.
With more focused discussions, solving specific issues is left to be discussed after the stand-up meeting.

One sole programmer expressed a strong negative feeling about the former stand-up meetings (every other day), going as far as saying he could feel humiliated.
His opinion about the new organization (every day) has improved in the sense that it is now neutral.

Two programmers said that when working on big tickets, telling every day the same thing is a bit redundant. 

\paragraph*{Impact of COVID-19}

The meetings are now held remotely, using a kanban board collaborative tool (Trello for the moment).
But programmers note that these virtual meetings are not as good, making informal communication more difficult and leading to ``\emph{a lack of motivation}''.
The Programmer 4 adds that ``\emph{we cannot know whether people are listening or not}''.
Also, ``\emph{at stand-up meetings where everyone has to talk, we sometimes feel a lack of motivation due to the isolation of the employees.
This has a [negative] impact on the productivity of the team.}''

\subsection{Team organizational change}
\label{sec:result-team}

\begin{figure}[htbp]
    \centering
    \includegraphics[width=.45\textwidth]{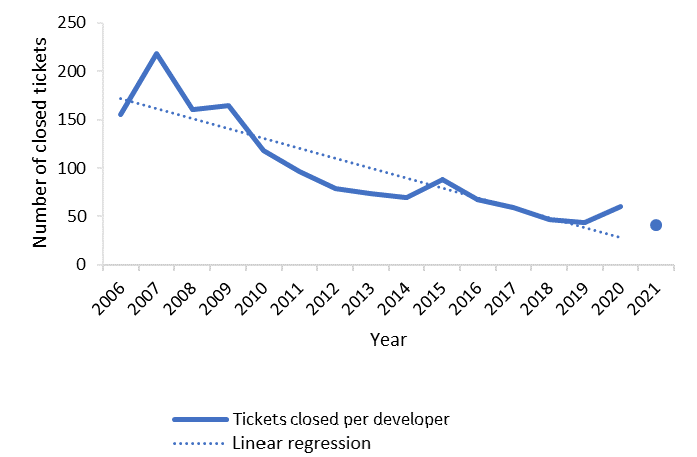}
   \caption{Number of closed tickets per programmer}
   \label{fig:closedTickets}
\end{figure}

\begin{figure}[htbp!]
    \centering 
    \includegraphics[width=.45\textwidth]{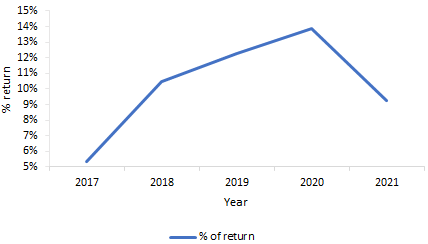}
    \caption{Return on closed tickets}
    \label{fig:ticketReturn}
\end{figure}

The new organization in business-oriented teams is generally  appreciated by  the participants.
The improved communication between programmers, testers, and analysts is perceived as speeding up the feedback loops between them thus making it easier for the programmers and testers to achieve their goals.
The Programmer 5 reports that ``\emph{Now we have better relationships with the analysts and testers.
The advantage is that the developers better understand the requirements and that the tester knows what to test.}''.
The testing manager added that ``\emph{We deliver faster to the customer.}''.
The perceived result is that tickets are closed faster and with more quality.
We tested this perception against actual data.
This new organization is only effective since the beginning of 2020, so it is too early to have strong data on its effects.
Figure \ref{fig:dashboard} (part (a)) does show a significant drop in the working time per ticket in 2020 and 2021, but 2019 was an especially bad year in this sense. 
The lifetime of tickets (Figure \ref{fig:dashboard}, part(b)) did drop in 2020, but it went slightly up again for the first four months of 2021.
These data do not provide a clear confirmation.
The average number of tickets closed per programmer, Figure \ref{fig:closedTickets}, does show a rise from a low point of 42 in 2019 to 60 in 2020.
Partial data on the first four months of 2021 (37 closed tickets per programmer) are also very encouraging.
Finally, Figure \ref{fig:ticketReturn} shows slightly higher percentage of return (rework) on closed ticket in 2020 (17.6\%) than in 2019 (14.4\%).
That number then drops on the first four months of 2021 (9.9\%).
This data in encouraging for the quality of the closed tickets.
These numbers cannot be attributed solely to the new team organization.
All new practices may have impacted the productivity of the programmers.

On the team new organization two problems were mentioned.
First, the analysts ended up with more work, now having to handle development cycle management (planning, monitoring).
Second, one of the two business teams sometimes ends up being the team for ``anything not for the first team'' and this makes it more difficult to manage.
Yet the general opinion remains that the advantages out-weight these two points.

\paragraph*{``Run team''} It also received positive and negative perceptions from the interviewees.

On the positive side, an analyst (not working on this team) expresses that it eases the planning of the development cycles (for the two other teams):
``\textit{We have an entry point which [\ldots] can handle [an urgent ticket] and it does not go to the business teams. Therefore, planning [of the business teams] is less impacted.}"
Two programmers feel that 
``\textit{[the run] team won its spurs [\ldots] we see a lot of tickets solved. Even if it's small tickets, we can't leave them out. And that satisfies the clients}."
Test management has a similarly good impression:
``\textit{programmers work more smoothly and have less change in priorities.}''
Finally, the Programmer 1 says that ``\textit{it allows one to vary the work, which is interesting}".

However, programmers that actually work on this team  mention the difficulty, the stress, or even ``\emph{a bit of a fear}", to be on it
(participation to the run team is rotating, see Section \ref{sec:chg-team}).
The main drawbacks mentioned are that (1) the rotation implies programmers may not always have the required competency to deal with an urgent ticket;
(2) when there is a ticket return, the original programmer may be back on his business team and another one needs to take the ticket and re-analyse it.
Such problems  rarely occur in the business teams which have more permanent personnel.



There are also difficulties with the management of this team:
For example, the customer manager 1 reports that ``\textit{the difficulty right now is that, as programmers rotate every three iterations, I don't necessarily have the same skill levels [available].
So, I can't give some tickets to some programmers due to a lack of skills.
This impacts my weekly goal.}''
Note, however, that although it makes life more difficult for programmers, this is intended by management as a strategy tending to global ownership of the code.

Among possible improvements, people suggest that this team should have permanent staffing (not rotating), but it is not clear whether they actually meant that they would prefer being permanently affected to their business team (which is seen as preferable).
Other programmers wished the ``run team''  would be organized as the two other ones, with formals analysis of the tickets, better planning of the development cycle.


\paragraph*{Impact of COVID-19}
With remote working, it can be difficult for the programmers to get in touch with the customer manager, and they sometimes end up having to do all the analysis work alone.
This is mostly felt for the run team where no formal analysis is done on tickets and the intended setup is that tickets are solved jointly between the programmer and the customer manager.
Remote working made this part more difficult.
Some programmers reports that``\emph{The benefits  of the ``run team'' is reduced because of slowness of communications due to the COVID-19.}''

\subsection{Software development cycle}
\label{sec:result-sprint}

The change in the development cycle (from cycles of three months to two weeks) is unanimously perceived as positive.
The benefits announced by interviewees are:
client requests are better managed;
clients and developers have a better visibility on the ongoing work and the feature to be delivered;
planning is better managed.




Despite all the positive aspects reported by participants, some difficulties or drawbacks were reported as well:
Time pressure is still strong, Programmer 3 and 4  regret that with short cycles there is no time for reflection or to apply good practices.
``\emph{The approach in cycles is not well-fitted for [the company's main product]. We have difficulties predicting how long a ticket will take}''.
According to the Programmer 3, this generates frustration ``\emph{in relation to the announced goals at the start of the cycle and what we actually achieve}''.
Another point is that deliveries are not tied to cycles, so programmers and analysts regret that ``\emph{we have no visibility on the release for which a ticket given should be treated}''.




\paragraph*{Impact of COVID-19}
Interviewees did not raise any special issue on development cycle due to remote working.

\subsection{Source code management}
\label{sec:result-svn}

The introduction of SVN  for source code management is almost unanimously perceived as a positive step.
``\emph{It's more reassuring}''.

But the change in working habits was difficult to put in place: 
``\textit{It generated frustration, some people were reluctant [\ldots] it did not ease the transition}''.

The main drawback was and remains conflict resolution.
As explained in Section \ref{sec:chg-svn}, PowerBuilder is not well suited for file based version control systems.

Branching management in SVN is not considered as advanced as in other version control system such as Git.
The current organisation is to have only one branch (per delivered version of the product).
This implies that all commits are pushed to the same branch and need sometimes to be later removed if the commit is not part of a delivery.
This causes extra manual work.
The solution envisioned to this problem is to switch to Git version control management soon.
This will allow isolating tickets in  specific branches and should ease the integration work.

\paragraph*{Impact of COVID-19}
Interviewees did not see any specific contribution of SVN to remote working.
It would seem natural that not having SVN would have made it more difficult to synchronize work between the programmers, but they explicitly stated that they did not see it that way.

\subsection{Code quality}
\label{sec:result-quality}

Upper management sees team code review (see Section \ref{sec:bck-quality}) as a way to achieve team ownership of the entire codebase (as opposed to author ownership on one's code).

All programmers do not seem to entirely share that understanding.
Newcomers do acknowledge this, a junior programmer regrets that \emph{``sometimes global code reviews are cancelled when there is an urgent delivery to ensure}'' and that ``\emph{there should be more of them}'' (weekly).
But more seasoned programmers propose to have less of them (monthly).
Their point is that it is difficult to find something interesting to show to the team every two weeks.
Programmers reporte that ``\emph{We don't always think to look piece code for code review when we are coding, because we are often in a hurry.}''
Even if a seasoned programmer acknowledges that it ``\emph{allowed [him] to reuse a function instead of recreating it}" on one occasion.


Another critic is that these reviews are oral and people forget the information shared after a while.
One programmer suggested keeping a written trace of these reviews.

Another proposal to renew the interest would be to organize the team code reviews by business domain or software quality domains (optimization, readability, ...), or to have exercise on which everybody would work together (coding dojo).

The CTO is happy with the current practice, feeling that ``\textit{The overall code quality is improving.}".

Another action was taken to improve code quality: an in-house ``linter'' for some rules (see Section \ref{sec:chg-quality}).
The linter runs on every commit and reports, by email, added and removed violations.
Since its installation at the beginning of 2021, 
178 commits resulted in an email with 48 positives (violations removed) and 130 negatives (violations added).
On the main development branch, there were 76 emails, and the number of rule violations went from 1113 to 669 (60\%).
The tool was well-received both by programmers and the development manager: ``\textit{These emails force to fix the reported errors little by little. Everyone does a small part, and it is less burdensome. The positive emails after fixing errors are good}".

One aspect of this good perception might be that the rules checked are in-house rules that are well known to all but still with many violations.
Past research \cite{Hora12b} showed that specific rules are better accepted than more generic ones.
The company is now considering switching to an external tool (Visual Expert\footnote{\url{https://www.visual-expert.com/}}) that would be more robust.
Whether it will be able to check the in-house rules should be an important point in the decision.

\paragraph*{Impact of COVID-19}
Here again, the main impact was that global code review meetings are held remotely.
This is generally seen as a drawback, but no special issue was raised during the interviews.

COVID-19 had no impact on the in-house ``linter", first because this action was initiated in 2021, and second, because it is fully automated (upon commit to the SVN repository) and communicates violations added or removed to the programmers by email.

\subsection{Dashboard of tickets}
\label{sec:result-dashboard}

The ``dashboard of tickets'' presents graphs similar to the Figure \ref{fig:dashboard}.
It was initially printed on a sheet of paper and exposed in the open-floor office.

People feel that there is a lack of communication around this tool.
Some added that having better access to the dashboard would improve their involvement in the new work processes as it will give constant feedback.
A developer proposed that the dashboard should generate a report every three development cycle (\ie six weeks),
and a report each year for evaluating the impact of the new practices on productivity.

A slightly negative feeling was from a developer saying that it is not a tool for developers but managers, and he does not see any utility in it.

\paragraph*{Impact of COVID-19}
As noted above, the dashboard used to be printed on a sheet of paper and exposed in the open-floor office.
When switching to remote work, no action was taken to make this dashboard available from the intranet.
That may be an explanation for the lack of communication that many noted.




\section{Threats to Validity}
\label{sec:threats}

This research is a case study aiming at raising questions for future research.
This has an impact on the kind of threats to validity we faced.
For example, we do not aim at offering conclusions that would be easily generalized to other contexts.

To discuss the possible threats, we followed the four perspectives of validity threats presented by \cite{Wohl12}: construct validity, internal validity, external validity and reliability.

\paragraph*{Construct validity}
Are we asking the right questions?
The questionnaire was built iteratively by two of the authors, one making successive proposals and the other commenting on it.
The questionnaire consists in a series of questions (see \tabref{questions}) asked successively for each identified point of interest (the practices).
The CTO interview differed because he has a very specific role, first as the main initiator of several of the new practices, second as he does not take part in the concrete development tasks.

\paragraph*{Internal validity}
Is there something inherent to how we collect and analyze the data that could skew our findings?
For case studies, this kind of threat is a bit different than usual because the goal is not to offer generalizable conclusions, but to raise points of interest for possible future research.

We  did aim to get a balanced view by including many different profiles in the interviews (CTO, development manager, customer manager, analyst, programmer, tester).
This covers all the stakeholder roles in the development of software in the company.
We do acknowledge that our participants are in majority programmers which could constitute a bias.
But this seems justified as it is a reflect of the actual demography of the company and many development teams.

The interviews were semi-structured with open questions to let participants come up with whatever comments they felt relevant.
We ensured participants of anonymity of their answers.
The interviews were conducted by the first author which is known by all participants as he has been doing research in the company for two years.

This author participated directly in the introduction of some of the practices reviewed.
He gave the 8 hours training on SVN, implemented the ticket dashboard and the in-house linter.
This could introduce a bias as people may have been wary of expressing negative opinions on these practices.
However, we estimated that the trust relationship established with the participants during these two years was enough to counter-balance the possible bias while being important to get more detailed answers.

We adopted the grounded theory approach to analyze the interviews and refine our analysis iteratively to lessen potential biases.
When possible, we validated the impressions of the participants (our findings) with empirical data.

\paragraph*{External validity}
Are our results generalizable for practices modernization?
Our data collection is limited to one case study and does not aim at being readily generalized.

\paragraph*{Reliability}
Can others replicate our results?
Our interview questions are available in \tabref{questions} and we describe our study design in Section \ref{sec:study-design}

\section{Conclusion}
\label{sec:conclusion}
This paper reports on an empirical study of the introduction of several practices recommended in agile development in a medium-sized company.
Our study consists of semi-structured interviews of 17 participants with different perspectives on the development process: CTO, customer manager, analyst, programmer, tester.
The changes coincided with the COVID-19 outbreak and the need to resort to remote working.
We attempt to empirically identify perceived benefits and drawback of agile practices introduced by the company.
We  also validated some of these perceptions with hard data.
Finally, we identified the impact of COVID-19 on these practices.

Our findings include:
\begin{itemize}
\item Introduction of a version management tool (SVN) is positively perceived, but the migration was difficult and required more efforts than the other new practices;

\item It must be noted that the technology used (PowerBuilder) creates extra difficulties as the source code is stored in a proprietary format and does not easily interact with, file oriented, version control managers;

\item Although unit testing has not yet been introduced, it is planned for the near future and some initial experiments (not reported here) were conducted.
The first impression is again that PowerBuilder will impose additional constraints with a difficulty to create viable context for the tests without launching a full blown application.
We conclude that there are obsolete technologies, still very much in use for legacy software, that makes it difficult to adopt up-to-date recommended practices.
Solutions adapted to these technologies must be proposed to help migrate to modern practices;

\item Although the daily stand-up meetings are perceived as a good practice, remote working had a negative impact on it and lowered its advantages;

\item It is important that daily stand-up meetings do occur daily. Every other day meetings do not have the same positive effects;

\item Shorter software development cycles and team organizational changes are positive even during the COVID-19;

\item The ``run team'' (specialized in answering short term issues) received mixed opinions.
It is acknowledged as facilitating planning but programmers do not like being part of it (membership is rotating);

\item Management would like to introduce some level of team ownership of the code as opposed to individual programmers owning their code.
Actions like the ``run team'' or team code reviews intend to achieve this.
But these actions are those least well-received by the developers.
We have no insight whether this is because they don't adhere to the long term goal, they don't see how the practices contribute to the goal, or some other reason;

\item A ``linter'' was introduced to email programmers when they commit code violating some in-house rule.
It was well-received by all.
It is hoped that it will ease the future introduction of unit testing and better continuous integration.
The positive reaction might be linked to the fact that it checks in-house rules as opposed to generic code quality rules \cite{Hora12b}.
\end{itemize}


\bibliographystyle{plain}
\bibliography{rmod,others,new}

\end{document}